\documentclass[%
reprint,
amsmath,amssymb,
prb,
]{revtex4-1}

\usepackage{graphicx}
\usepackage{dcolumn}
\usepackage{bm}
\usepackage{color}

\begin{document}

\preprint{APS/123-QED}

\title{\textit{Ab initio} study of the ferroelectric strain dependence and 180$^\circ$ domain walls in the barium metal fluorides BaMgF$_4$ and BaZnF$_4$}

\author{Maribel N\'u\~nez Valdez}
\email{nunez\_valdez.m@mipt.ru}
\affiliation{Materials Theory, ETH Z\"urich, Wolfgang-Pauli-Strasse 27, CH-8093 Z\"urich, Switzerland\\ Moscow Institute of Physics and Technology, Dolgoprudny, Moscow Region, Russia}
\author{Hendrik Th.\ Spanke}
\author{Nicola A.\ Spaldin}
\affiliation{%
Materials Theory, ETH Z\"urich, Wolfgang-Pauli-Strasse 27, CH-8093 Z\"urich, Switzerland
}%

\date{\today}

\begin{abstract}
We investigate the strain dependence of the ferroelectric polarization and the structure of the ferroelectric domain walls in the layered 
perovskite-related barium fluorides, BaMF$_4$ (M=Mg, Zn). 
The unusual ``geometric ferroelectricity'' in these materials is driven by the softening of a single polar phonon mode consisting of 
rotations of the MF$_6$ octahedra accompanied by polar displacements of the Ba cations, and in contrast to conventional ferroelectrics
involves minimal electronic rehybridization. 
We therefore anticipate a different strain dependence of the polarization, and alternative domain wall structures compared with those 
found in conventional ferroelectric materials. 
Using first-principles calculations based on density functional theory (DFT) within the general gradient approximation (GGA), we calculate
the variation of the crystal structure and the ferroelectric polarization under both compressive and tensile strain.  
We perform structural relaxations of neutral domain walls between oppositely oriented directions of the ferroelectric 
polarization and calculate their corresponding energies to determine which are most likely to form. We compare our results to literature
values for conventional perovskite oxides to provide a source of comparison for understanding the ferroelectric properties of alternative 
non-oxide materials such as the barium fluorides.   
\end{abstract}

\pacs{Valid PACS appear here}
\maketitle

\section{Introduction}
Ferroelectric and multiferroic oxides are widely studied because of their fundamental interest and for technological applications 
such as non-volatile random access memories \cite{auciello}, piezoelectric actuators and sensors \cite{muralt}, pyroelectric 
detectors \cite{zhang}, and electro-optic and non-linear optical devices \cite{wessels}. Indeed, it is often assumed that the
presence of oxygen, which forms highly polarizable bonds with transition metal cations, is a requirement for good ferroelectric
behavior. Recently, however, research on ferroelectric materials based on alternative chemistries without oxygen has received 
renewed interest. In particular, the class of barium metal fluorides, BaMF$_4$ (for a review see Ref.~\onlinecite{scott1979} and
references therein), such as BaMgF$_4$ \cite{buchter} could prove to be important because
of the wide band gaps and associated transparency of fluorine-based compounds, which makes them attractive for advanced photonic 
and optoelectronic applications \cite{meyn}.  

\begin{figure}[h]
\includegraphics[width=0.5\textwidth]{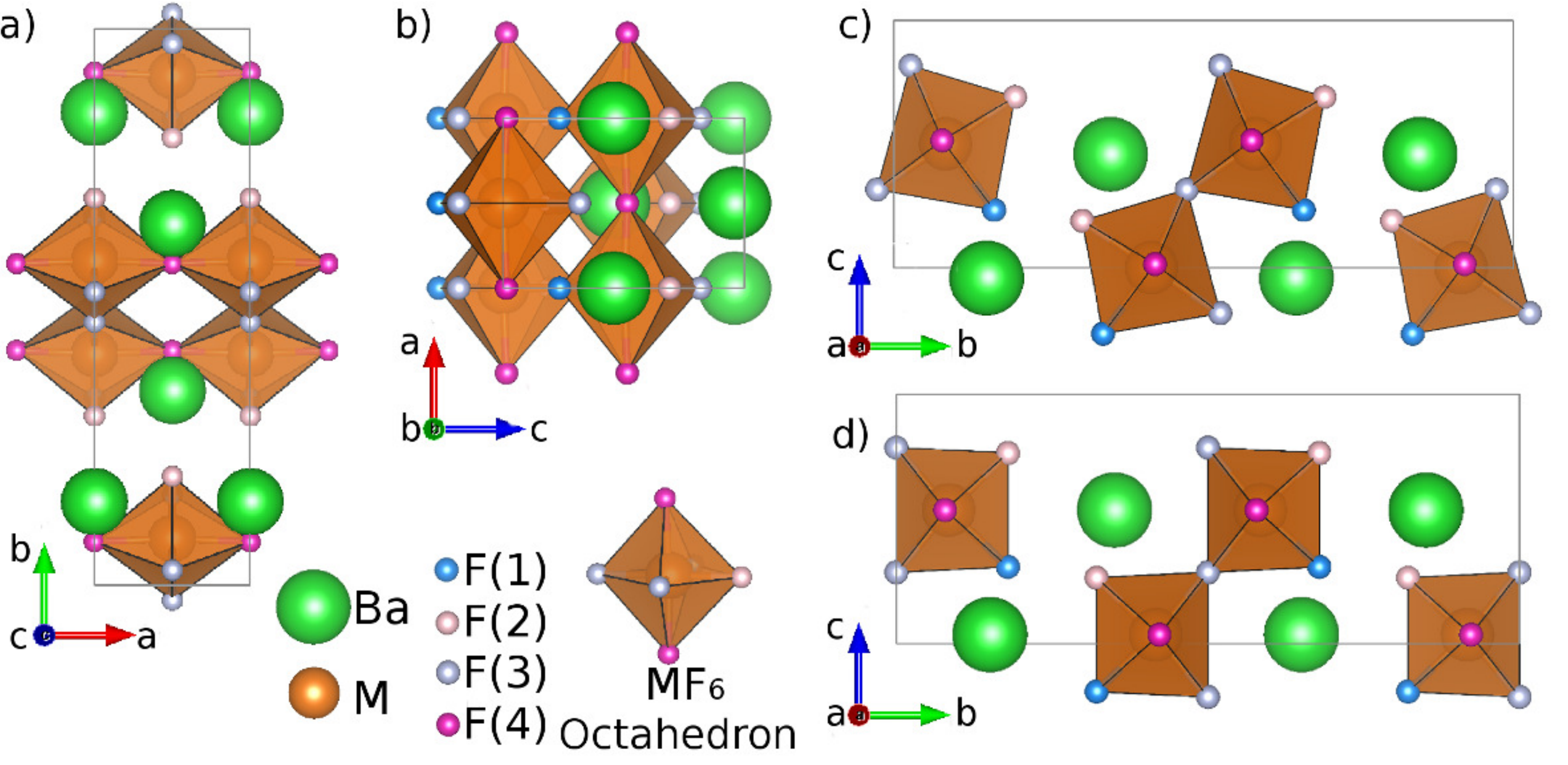}
\caption{\label{structure} (Color online) Structure of BaMgF$_4$. a-c) Ferroelectric $Cmc2_1$ phase in different
orientations. The calculated lowest energy structure, which is similar to the experimental one, is shown. d) Hypothetical $Cmcm$ 
centrosymmetric reference phase. The ferroelectric and centrosymmetric structures of BaZnF$_4$ are qualitatively
similar to those of BaMgF$_4$. The conventional primitive unit cell, which contains 24 atoms (four formula units), is indicated in gray. (All crystal visualizations in this paper were performed using VESTA\cite{vesta}).}
\end{figure}

The BaMF$_4$ barium metal fluorides form in a bilayered, perovskite-related base-centered orthorhombic structure 
(space group No. 36, $Cmc2_1$) as shown in Fig.~\ref{structure}. The divalent M cation can be a $3d$ transition metal ion (Mn, Fe, Co or Ni) 
or a non-magnetic divalent ion (Mg or Zn), and is octahedrally coordinated by fluorine anions. Two-layer slabs of corner-sharing MF$_6$ 
octahedra lie perpendicular to the crystallographic $b$-axis with Ba cations in planes between the slabs \cite{schnering,eibschutz}. 
The structure is polar, and ferroelectric switching using a pulsed-field technique has been demonstrated at room temperature for all 
members of the BaMF$_4$ family except for M = Mn and Fe \cite{eibschutz,scott1979}. The high-temperature paraelectric reference phase has not been
identified experimentally, because melting occurs before the ferroelectric Curie temperature, $T_C$ (estimated to be between 1100 and 1600
K by extrapolating the temperature-dependent dielectric constants \cite{didomenico}) is reached. First-principles electronic structure
calculations have shown, however, that the ferroelectric ground state can be reached from a prototypical centrosymmetric $Cmcm$ structure
[Fig.~\ref{structure} (d)] via a single polar phonon mode which consists of a rotation of the MF$_6$ octahedra accompanied by a displacement of the Ba$^{2+}$ atoms along the $b$ axis \cite{ederer}. The driving force for this so-called {\it geometric ferroelectricity} is the combination of size effects and the layered geometric coordination of the crystal lattice, rather than the usual electronic rehybridization found in conventional ferroelectrics.
As a result, the Born effective charges, which reflect the degree of rehybridization during a polar distortion, are close to the formal
ionic charges in contrast to the anomalously large values characteristic of conventional ferroelectrics. The same mechanism is believed to
occur in the layered, perovskite-related rare-earth titanates R$_2$Ti$_2$O$_7$ \cite{lopez,frank}
and a related improper version occurs in the hexagonal rare-earth manganites \cite{vanAken_et_al:2004,Fennie}.

We anticipate that this unconventional mechanism for the ferroelectric polarization in the BaMF$_4$ family might lead to quite
different behavior in two properties that are particularly relevant for the incorporation of ferroelectrics into thin film
devices. The first is the strain dependence of the ferroelectric polarization, which is particularly important when ferroelectrics are grown on
substrates with mismatched lattice constants. The polarization-lattice coupling is substantial in many conventional oxide ferroelectrics 
\cite{ederer2}, and is believed to be driven by the large electronic rehybridizations reflected in the anomalous
Born effective charges, therefore we expect smaller effects here. The second is the nature of the domain walls separating regions of 
opposite polarization. These have been studied extensively both theoretically and experimentally in 
conventional ferroelectric perovskite oxides because they have a profound influence on the material physical properties, in
particular the ferroelectric hysteresis. The identity and structure of the lowest energy domain walls in conventional perovskite
oxides are now well established
(see for example Refs. [\onlinecite{Stemmer_et_al:1995, Streiffer_et_al:1998, Padilla/Zhong/Vanderbilt:1996, Meyer/Vanderbilt:2002}]).
In addition, the interaction of domain walls with point defects such as oxygen vacancies, and the resulting effects on the properties are topics
of tremendous current interest \cite{Becher_et_al:2015}.
With the unusual rotational mechanism for ferroelectricity in BaMF$_4$, the domain walls might more closely resemble anti-phase
boundaries, with different energetics and thicknesses from their conventional ferroelectric counterparts.
To the best of our knowledge, however, calculations exploring the strain dependence of polarization and the structure and properties 
of the domain walls in BaMF$_4$ ferroelectrics have not been performed; this is the goal of this work. 

The remainder of this paper is organized as follows: In section II we describe the technical details of our calculations. 
Section III, the main part of the article, contains our results for our two representative materials, BaMgF$_4$ and 
BaZnF$4$. Specifically, we present the structures of the low-energy neutral domain walls obtained using structural optimizations 
of atomic positions, and the dependence of the spontaneous polarization on the strain. 
In section IV we summarize our main findings and present our conclusions.  

\section{Computational Details}

Our calculations were performed using the Vienna \textit{Ab initio} Simulation Package (VASP)\cite{kresse96b} within the 
projector-augmented plane wave (PAW)\cite{blochl,kresse99} method of density functional theory (DFT)\cite{dft1,dft2}.  
We used the general-gradient approximation (GGA) in the prescription by Perdew, Burke and Ernzerhof (PBE)\cite{perdew2} 
for the exchange-correlation potential. We used the default PAW potentials with the valence electronic configurations $5s^25p^66s^2$
for Ba, $3d^{10}4s^2$ for Zn, $3s^2$ for Mg, and $2s^22p^5$ for F. A plane-wave cutoff energy of  500 eV  and a Brillouin-zone 
$k$-point sampling of 6x4x6 within the 24-atom unit cell were used. Convergence was assumed when the forces on each atom were
smaller than 1 meV/\AA~ and the total energy changes less than 10$^{-8}$ eV. The electronic contributions to the spontaneous polarization 
($P_S$), defined as the difference in polarization between the ferroelectric ground state structure ($Cmc2_1$) and the postulated high 
symmetry paraelectric phase ($Cmcm$), were calculated using the Berry phase approach \cite{berry1,berry2,berry3} by integrating over 
six homogeneously distributed k-point strings, parallel to the reciprocal crystallographic $c$-axis, each containing ten $k$-points.

\begin{table}[h]%
\caption{\label{structure_par}%
Our calculated structural parameters [*] at zero temperature for the $Cmc2_1$ ferroelectric phases of BaMgF$_4$ and BaZnF$_4$. Experimental data at 10 K from Ref. [\onlinecite{posse}] for both materials, and DFT results from Ref. [\onlinecite{cao}] for BaZnF$_4$ are shown for comparison. All atomic positions have Wyckoff symmetry $4a$.}
\begin{ruledtabular}
\begin{tabular}{rddddd}
\textrm{Para-}&
\multicolumn{2}{c}{\textrm{Mg}}&
\multicolumn{3}{c}{\textrm{Zn}}\\
\textrm{meter}&
\multicolumn{1}{c}{\textrm{DFT}}&
\multicolumn{1}{c}{\textrm{EXP}}&
\multicolumn{2}{c}{\textrm{DFT}}&
\multicolumn{1}{c}{\textrm{EXP}}\\
 &
\multicolumn{1}{c}{[*]} &
\multicolumn{1}{c}{\textrm{Ref. [\onlinecite{posse}]}} &
\multicolumn{1}{c}{[*]} &
\multicolumn{1}{c}{\textrm{Ref. [\onlinecite{cao}]}} &
\multicolumn{1}{c}{\textrm{Ref. [\onlinecite{posse}]}} \\
\colrule
 $a_0$ (\AA) & 4.16 & 4.119 & 4.25 & 4.281 & 4.191 \\
 $b_0$ (\AA) & 14.83 & 14.463 & 14.88 & 14.700 & 14.513 \\
 $c_0$ (\AA) & 5.93 & 5.812  & 5.97 & 5.921 & 5.835 \\
Ba\hspace{.3cm} x & 0.5 & 0.5 & 0.5 & 0.5 & 0.5 \\
   y & 0.350 & 0.351 & 0.351 & 0.3520 & 0.352 \\
   z & 0.460 & 0.536 & 0.455 & 0.4575 & 0.537 \\
M\hspace{.35cm} x & 0.0 & 0.0 & 0.0 & 0.0 & 0.0 \\
   y & 0.416 & 0.414 & 0.414 & 0.413 &0.413 \\
   z & 0.002 & 0.0 & -0.002 & 0.0 & 0.0 \\
F(1) x & 0.0 & 0.0 & 0.0 & 0.0 & 0.0 \\
   y & 0.338 & 0.306 & 0.335 & 0.333 & 0.303 \\ 
   z & 0.734 & 0.803 & 0.726& 0.727 & 0.800 \\
F(2) x & 0.0 & 0.0 & 0.0 & 0.0 & 0.0 \\
   y & 0.304 & 0.334 & 0.302 & 0.301 & 0.330 \\
   z & 0.191 & 0.261 & 0.193 & 0.198 & 0.262\\
F(3) x & 0.0 & 0.0 & 0.0 & 0.0 & 0.0 \\
   y & 0.527 & 0.473 & 0.531 & 0.531 & 0.471 \\
   z & 0.817 & 0.692 & 0.831 & 0.830 & 0.673\\
F(4) x & 0.5 & 0.5 & 0.5 & 0.5 & 0.5 \\
   y & 0.422 & 0.422 & 0.423 & 0.423 & 0.422 \\
   z & 0.015 & -0.010& 0.017 & 0.017 & 0.983\\
\end{tabular}
\end{ruledtabular}
\end{table}


\section{Results}

\subsection{Structural, Electronic, and Ferroelectric Properties}

We begin by calculating the lowest energy structures and lattice parameters for the bulk ferroelectric ($Cmc2_1$) phases of BaMgF$_4$ 
and BaZnF$_4$. Our 0 K results, shown in Table \ref{structure_par}, compare reasonably with experimental measurements at $\sim10$ K extracted from synchrotron powder diffraction data \cite{posse} and a previous DFT calculation for BaZnF$_4$ \cite{cao}. Our calculated atomic positions are in good agreement 
with the experimentally determined positions along the $a$ and $b$ directions, with larger deviations in the $c$ direction. Likewise our $a$ and
$c$ lattice parameters are close to the measured values, with a difference of $\sim2.5$\% for the $b$ lattice parameter perpendicular to the
layers, likely due to the GGA overestimating the weak bonding between the layers. 

\begin{figure}[h]
\includegraphics[width=0.49\textwidth]{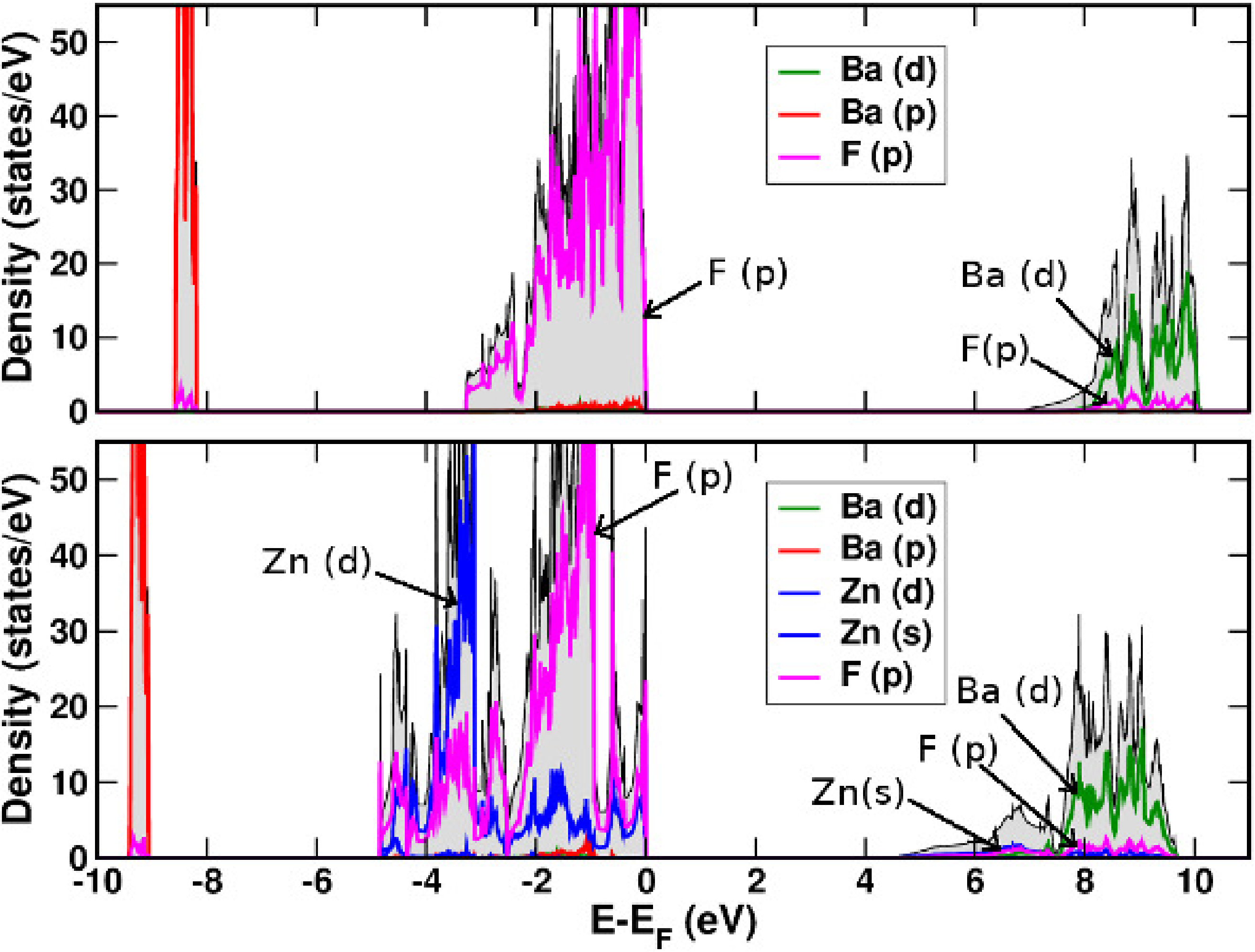}
\caption{\label{dos_BaMF4} (Color online) Total (black line and gray shaded) and partial (color) densities of states of bulk  a) BaMgF$_4$ and b) BaZnF$_4$. Ba $p$ states are shown in red, Ba $d$ states in green, and F $p$ states in magenta. Zn $d$ and $s$ states are in blue. } 
\end{figure}

In Fig.~\ref{dos_BaMF4} we show our calculated densities of states. We see that both compounds are strongly insulating with large DFT band 
gaps (6.9 and 4.5 eV for BaMgF$_4$ and BaZnF$_4$ respectively). The top of the valence bands is formed primarily from F $2p$ states and the 
lower part of the conduction bands from Ba 5$d$ states, with negligible hybridization between them. A notable difference between the two materials
is the presence of Zn $3d$ states mixed with the F $2p$ states at the bottom of the valence band in BaZnF$_4$, and Zn $4s$ states at the bottom
of the conduction band leading to the smaller gap in this case. Mg $s$ states are minimally present in the range shown. 

\begin{figure}[h]
\includegraphics[width=0.49\textwidth]{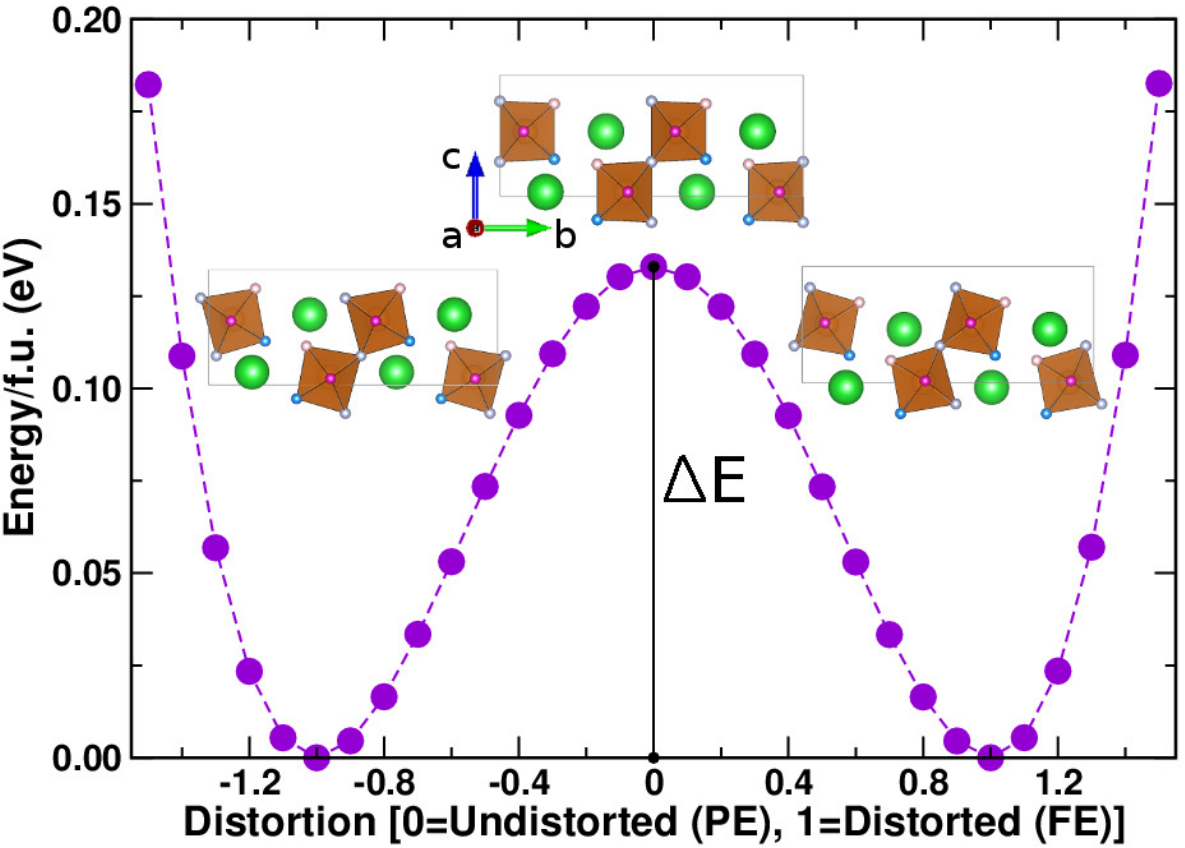}
\caption{\label{doublewell} (Color online) Energy per formula unit (f.u.) of BaMgF$_4$ as a function of the magnitude of the ferroelectric structural
distortion. For BaZnF$_4$ the form of the double-well potential is qualitatively similar but the energy barrier, $\Delta E$, is larger. The left and
right insets show the MgF$_6$ octahedral rotation patterns in the two ferroic ground states compared to the unrotated PE phase (central inset).}
\end{figure}

Fig.~\ref{doublewell} shows the dependence of the total energy per formula unit (f.u.) on the pattern of atomic displacements that transforms
the paraelectric (PE) structure to the ground state ferroelectric (FE) structure for BaMgF$_4$. We see the usual double-well potential 
characteristic of proper ferroelectrics with an energy barrier, $\Delta E$, between the two equivalent ferroic ground states of 0.133 eV/f.u.; 
the corresponding barrier for BaZnF$_4$ is 0.218 eV/f.u. For comparison, in the magnetic members of this fluoride family, with M = Mn, Fe, Co, 
and Ni, $\Delta E$ ranges from $\sim0.025$ to $\sim0.2$ eV/f.u.\cite{ederer}; the conventional oxide perovskite ferroelectrics BaTiO$_3$ and 
PbTiO$_3$ have energy barriers of .018 and .200 eV/f.u. respectively\cite{cohen}. 

Our calculated spontaneous polarizations, $P_S$, obtained from the difference in polarization between the undistorted $Cmcm$ PE phase and the 
$Cmc2_1$ FE ground state along the same branch of the polarization lattice are shown in Table~\ref{bulkPs}. We see that the values obtained 
using the Berry phase approach are similar to those obtained from multiplying the displacements of the ions with their formal ionic point 
charges (Ba$^{2+}$, M$^{2+}$, F$^{-}$), indicating 
that the Born effective charges, $Z^*$, are close to their formal values and that the ferroelectric mechanism is of geometric nature with 
no significant charge transfer between cations and anions \cite{ederer}.
As stated above, we expect that these non-anomalous Born effective charges might lead to a different strain-polarization coupling from that found in
conventional ferroelectrics and we investigate this next.

\begin{table}[h]
\caption{\label{bulkPs}%
Spontaneous polarizations calculated by summing over the product of the formal charges times the displacements, using the Berry phase approach, and measured
experimentally ($\ddagger$ Ref. [\onlinecite{didomenico}], $\dagger$ Ref. [\onlinecite{kannan}], $\star$ Ref. [\onlinecite{villora}])}
\begin{ruledtabular}
\begin{tabular}{cdd}
\textrm{$P_S$}&
\multicolumn{1}{c}{\textrm{BaMgF$_4$}}&
\multicolumn{1}{c}{\textrm{BaZnF$_4$}}\\
\textrm{(001)}&
\multicolumn{1}{c}{\textrm{$\mu$C/cm$^2$}}&
\multicolumn{1}{c}{\textrm{$\mu$C/cm$^2$}}\\
\colrule
\textrm{Formal charges }&
8.9 & 11.4 \\
\textrm{Berry phase}&
10.1 & 13.2\\
\textrm{Experimental}&
 7.7^\ddagger & 9.7^\ddagger \\
 & 6.9^\dagger  & 9.0^\star
\end{tabular}
\end{ruledtabular}
\end{table}

\subsection{Effect of strain on the ferroelectric polarization}
The thin film geometry, in which a $\sim$nm-thick layer of ferroelectric material is grown on a substrate or metallic electrode, 
is important in device architectures, and can be used to modify the ferroelectric behavior through strain induced via coherent 
heteroepitaxy with the substrate. In conventional perovskite ferroelectrics such strain-polarization coupling can be strong, leading
for example, to the onset of ferroelectricity in otherwise paraelectric SrTiO$_3$ \cite{Haeni_et_al:2004} and the enhancement of the polarization
and coercivity in ferroelectric BaTiO$_3$ \cite{Choi_et_al:2004}. First-principles studies \cite{ederer2} have rationalized the magnitude
of the strain dependence in terms of the material's piezoelectric and elastic constants, which in turn are often large in oxide ferroelectrics. 
Motivated by these features, and by the different nature of the ferroelectric polarization in the fluoride compounds, we now calculate the
strain dependence of the ferroelectric polarization for BaMgF$_4$ and  BaZnF$_4$.  

\begin{figure}[h]
\includegraphics[width=0.49\textwidth]{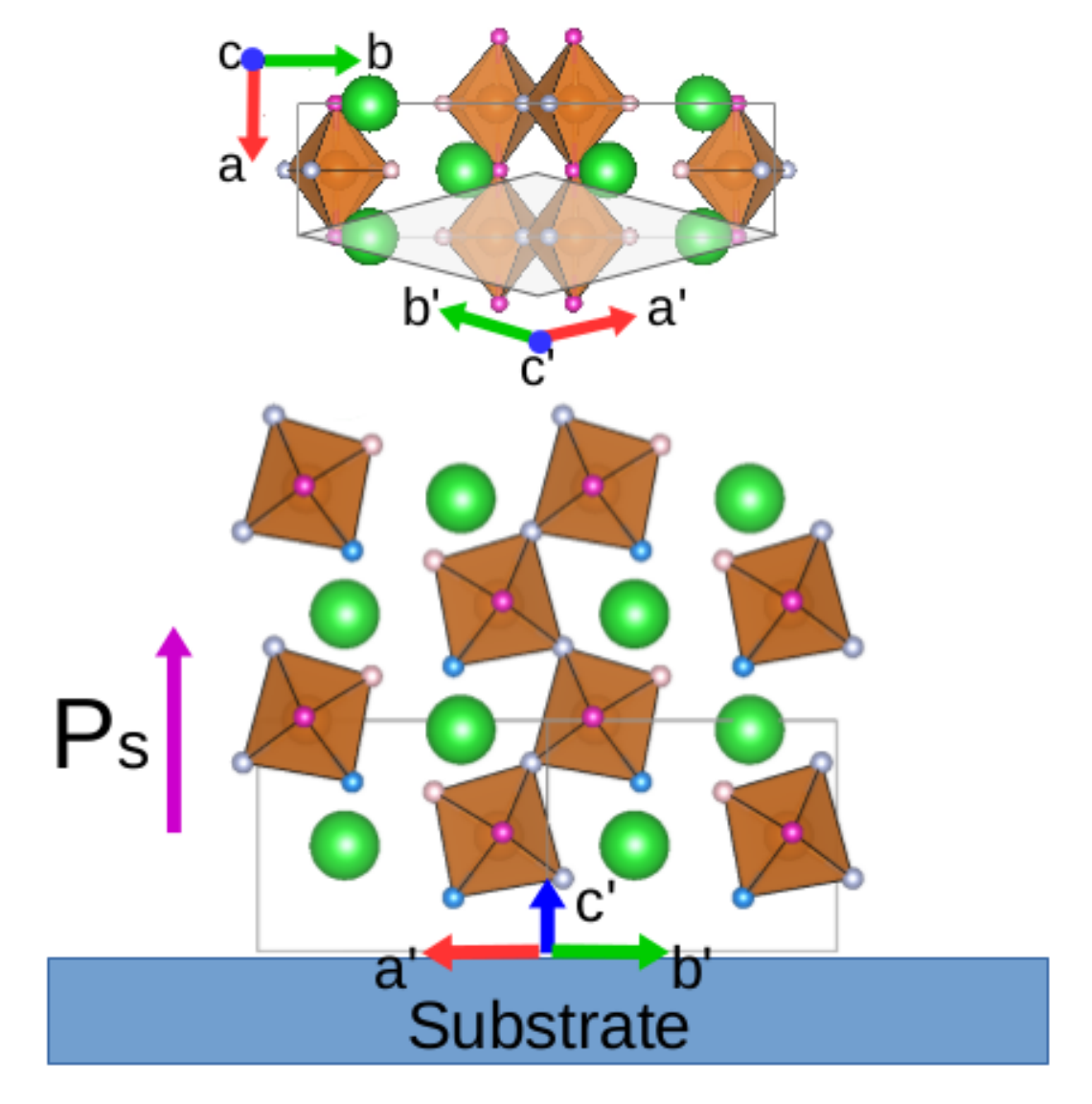}%
\caption{\label{BaMgF4_and_substrate} Orientation of BaMF$_4$ relative to the substrate adopted in this work. The 12 atom primitive cell is indicated and compared to the 24 atom cell (top figure).}
\end{figure}

For our strain calculations, we use a 12-atom primitive cell, which is connected to the conventional unit cell through the 
relationships $\vec{a}'=\frac{1}{2}(a,-b,2c)$, $\vec{b}'=\frac{1}{2}(a,b,2c)$ and $\vec{c}'=\vec{c}$, see Fig. \ref{BaMgF4_and_substrate}. This choice of system of reference is convenient because in this set up $|\vec{a}'|=|\vec{b}'|$. Strain is generated by fixing the lattice parameters corresponding to the lateral directions of the substrate ($a'$ and $b'$), relaxing 
the internal ionic degrees of freedom, and determining the out-of-plane lattice parameter ($\vec{c}'$) by means of an equation 
of state. We induce compressive and tensile strains between -3\% and +3\%, where the misfit strain is defined as $\varepsilon=\frac{|\vec{a}'|}{|\vec{a}'_0|}-1=\frac{|\vec{b}'|}{|\vec{b}'_0|}-1$. 
Note that in this orientation the ferroelectric polarization lies perpendicular to the plane of the film, a geometry that
is desirable for device applications but in practice might be difficult to achieve through conventional layer-by-layer growth 
methods.

\begin{figure}[h]
\includegraphics[width=0.49\textwidth]{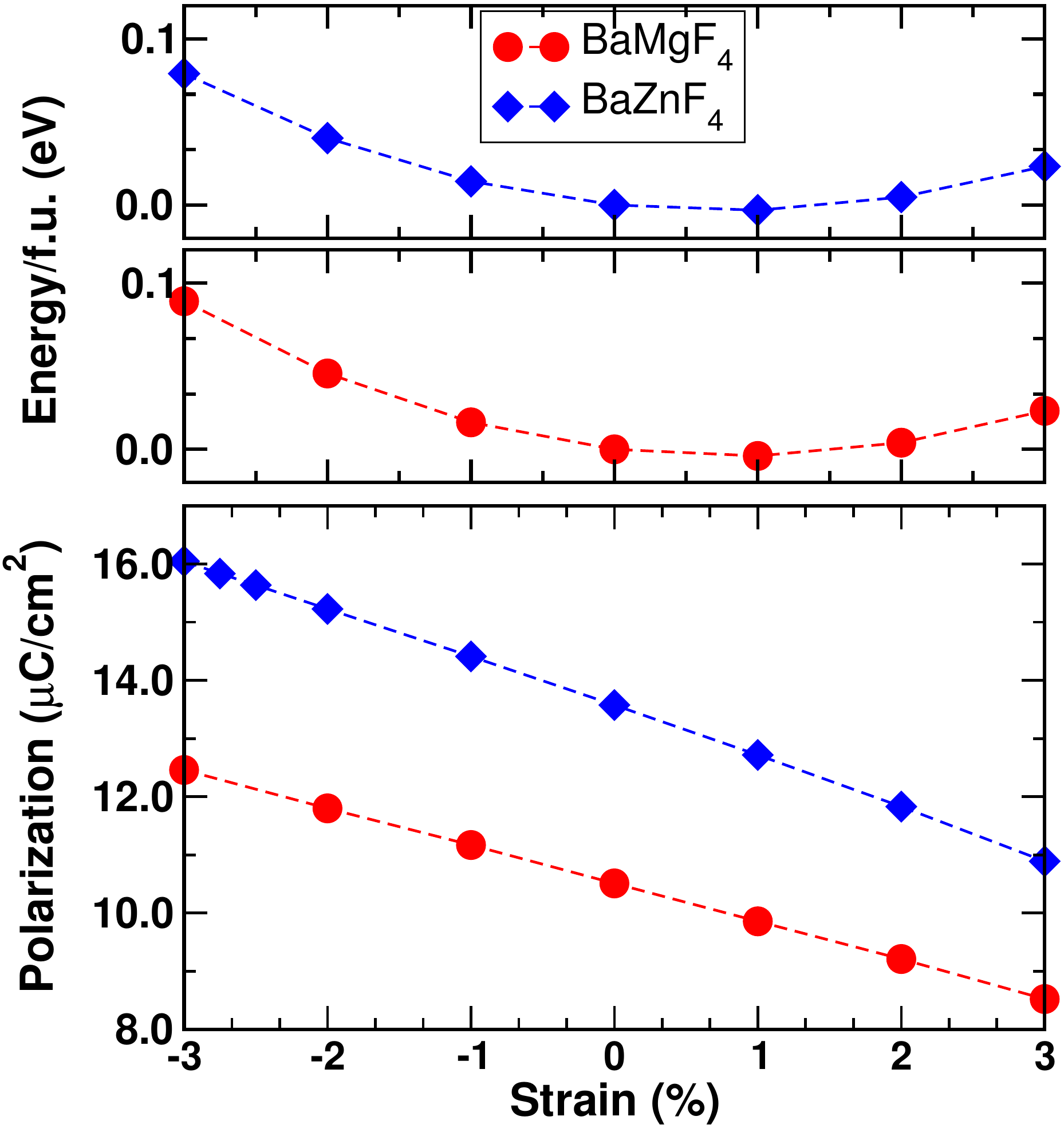}
\caption{\label{pol_strain} (Color online) a) Energy per formula unit (f.u.) and b) spontaneous polarization $P_S$ as a function of strain ($\epsilon$) calculated using the Berry phase approach for BaMF$_4$ (M=Mg, Zn).}
\end{figure}

Our calculated internal energies and polarizations as a function of strain are shown in Fig.~\ref{pol_strain}. The energy variations over the $\pm$ 3\% strain
range are less than 0.3\% for compressive strain and 0.1\% for tensile strain with respect to the zero-strain energy (Fig.~\ref{pol_strain} a). Then, we note that the polarization direction remains along the out-of-plane crystallographic $c$-axis for all strain values studied, in contrast to many perovskite oxides in which the polarization becomes in plane for tensile strain. Indeed, the spontaneous polarization varies close to linearly with strain for the
entire range considered, with variations in magnitude from around +20\% to -15\%  for both compounds compared to their unstrained
values. While less dramatic than in some
oxide counterparts, these values are not insignificant and should not be ignored in creating heterostructures with lattice-mismatched
materials. This nearly linear response of the spontaneous polarization to strain, $\epsilon$, indicates that the barium fluorides also satisfy the 
relationship discussed for conventional ferroelectrics in Ref. \onlinecite{ederer2} that
\begin{equation}
\Delta P=\left(2c_{31}-\frac{c_{33}}{n}\right)\epsilon=c_{\textrm{eff}}\epsilon \quad .
\end{equation}
Here $\Delta P$ is the change in polarization, $c_{31}$ and $c_{33}$ are components of the piezolectric tensor, and $n$ is the Poisson ratio. 
Our effective piezoelectric constants, $c_{\textrm{eff}}$, are $-65\mu$C/cm$^2$ and $-86\mu$C/cm$^2$ for BaMgF$_4$ and BaZnF$_4$, respectively,
comparable to that of rhombohedral BiFeO$_3$($R3c$) (-85$\mu$C/cm$^2$) but an order of magnitude smaller than those of BaTiO$_3$ and PbTiO$_3$. 
Our calculated electronic band structures (not shown) indicate minimal change in band gap with strain.

\subsection{Ferroelectric Domains: Formation of 180$^\circ$ Domain Walls}\label{domain wallsec}

\begin{figure}[h]
\includegraphics[width=0.49\textwidth]{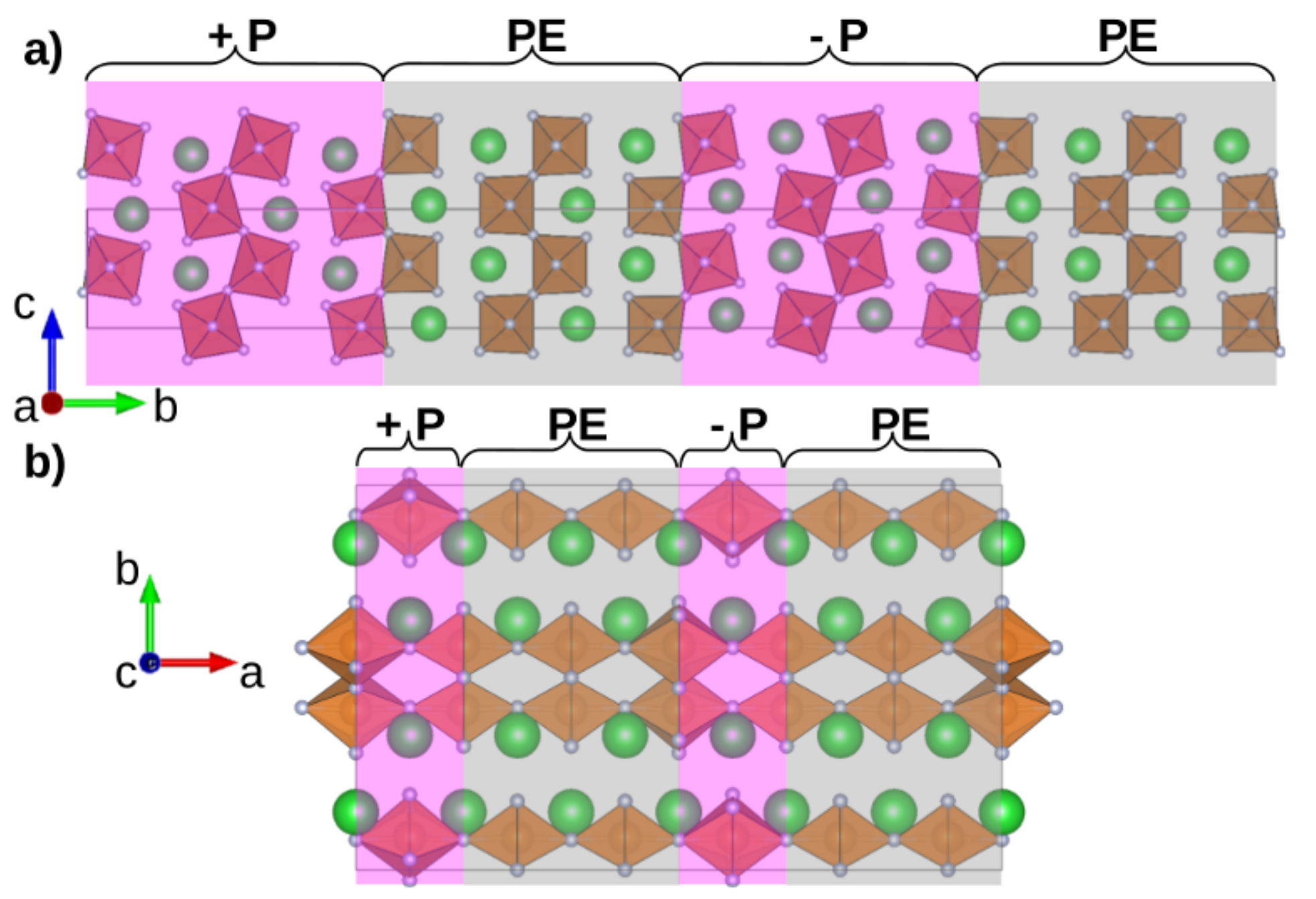}
\caption{\label{dwconfig} Starting a) (101)-DW and b) (011)-DW configurations for BaMgF$_4$ (for BaZnF$_4$ the setup is analogous). 
We show two supercells along the $c$ direction in a) only to help with visualization. Ions in PE regions (including the ones 
directly at the boundaries) are allowed to relax from the initial PE positions. The other regions have polarizations fixed 
along $+P$ and $-P$ respectively.}
\end{figure}

The domain walls between regions of differently oriented polarization in ferroelectrics are known to influence the 
ferroelectric switching behavior as well as to have functional properties in their own right \cite{Seidel_et_al:2009,Meier-et-al:2012}.
Much is known about the structure and energetics of domain walls in perovskite oxide ferroelectrics, both from first-principles density 
functional calculations (see for example Refs.~[\onlinecite{Meyer/Vanderbilt:2002,Lubk/Gemming/Spaldin:2009,kumagai,dieguez,ren}]) and from detailed experimental studies using for example high-resolution
transmission electron microscopy (see for example Refs.~[\onlinecite{jia}] and [\onlinecite{zhang2012}]). However information about ferroelectric domain walls in the BaMF$_4$ compounds is to our knowledge completely lacking; we provide the first calculations here.

In most domain walls, the component of polarization perpendicular to the wall is constant, so that
\begin{equation}
(\mathbf{P}_{\mbox{A}}-\mathbf{P}_{\mbox{B}})\cdot\mathbf{n}=0,
\end{equation}
where $\mathbf{P}_{\mbox{A}}$ and $\mathbf{P}_{\mbox{B}}$ are the spontaneous polarizations of the two domains. This condition avoids
a divergence of the electrostatic potential which would require a screening by additional charges and so such walls are called neutral 
(as opposed to charged) domain walls. In this work we restrict our discussion to neutral domain walls. In addition, we consider only 
180$^\circ$ domain walls, in which the orientation of the polarization changes by 180$^\circ$ across the wall, and leave for other 
orientations such as 90$^\circ$ domain walls for future investigation. We explore two geometries, with the 
normal vector $\mathbf{n}$ parallel to the crystallographic $a$- and $b$-axes in turn. To calculate the domain wall structures and
energetics we construct supercells of the form 1x4x1 times the primitive unit cell (containing 96 atoms) and 6x1x1 times the primitive
unit cell (144 atoms) for walls parallel to the $ac$- and $bc$-planes; we refer to these hereafter as (101)-DWs and (011)-DWs respectively. 
Within each supercell we impose two oppositely oriented domains with polarization parallel and antiparallel to the $c$ axis and two domain 
walls. The central slabs of each domain are constrained to their calculated bulk ferroelectric structures (see Fig.~\ref{dwconfig}). We 
then relax the atoms in the wall regions to their lowest energy configurations using the same convergence criteria as in Section II. 
The energy of a domain wall is then given by
\begin{equation}
E_{\textrm{domain wall}}=\frac{E-E_0}{2S},
\end{equation}
where $E$ is the total energy of the supercell configuration in the presence of domain walls, $E_0$ is the reference energy of bulk 
BaMF$_4$ (computed for the same corresponding supercell), and $S$ is the area of the domain wall (of which there are two per supercell). 
The convergence of the domain wall energies with respect to supercell size was tested by adopting different sizes with 72 and 96 atoms 
for the (101)-, and 96, 120, and 144 atoms for the (011)-DW configurations. The (011)-DWs converged more slowly and required larger 
supercells due to the corner sharing of the octahedra perpendicular to the domain wall.
Table \ref{Edomain wall} shows our calculated energies for the two domain wall configurations investigated here as well as literature
values for other selected ferroelectrics. 

\begin{table}[h]
\caption{\label{Edomain wall}%
Domain wall energies calculated in this work for the (101)-DW ($\dagger$) and  (011)-DW ($\ddagger$) configurations of BaMgF$_4$ and BaZnF$_4$, 
as well as the oxide ferroelectrics BiFeO$_3$, PbTiO$_3$ and hexagonal YMnO$_3$. The measured Curie temperatures and calculated polarizations
are also shown for comparison. (References: $\star$-This work, $*$-[\onlinecite{didomenico}], $a$-[\onlinecite{wang}], $b$-[\onlinecite{lubk,dieguez}], $c$-[\onlinecite{jaffe}], $d$-[\onlinecite{meyer}], $e$-[\onlinecite{gibbs}], $f$-[\onlinecite{fujimura}], $g$-[\onlinecite{kumagai}]).}
\begin{ruledtabular}
  \begin{tabular}{cccc}
    \textrm{Material} & $T_C$ \textrm{(K)}&
    $P_S$ \textrm{($\mu$C/cm$^2$)} & \textrm{E$_\textrm{domain wall}$ (mJ/m$^2$)}\\
    \colrule
    \textrm{BaMgF$_4$} & \textrm{1263$^*$} & \textrm{10.1} & \textrm{72$^\dagger$} \\
    \multicolumn{3}{c}{} & \textrm{148$^\ddagger$} \\
    \textrm{BaZnF$_4$} & \textrm{1083$^*$} & \textrm{13.2} & \textrm{185$^\dagger$} \\
    \multicolumn{3}{c}{} & \textrm{159$^\ddagger$} \\
    \textrm{BiFeO$_3$} & \textrm{1103$_a$} & \textrm{90$_a$} & \textrm{80-800$_b$} \\
    \textrm{PbTiO$_3$} & \textrm{765$_c$} & \textrm{75$_c$} & \textrm{132$_d$} \\
    \textrm{h-YMnO$_3$} & \textrm{1258$_e$} & \textrm{5.6$_f$} & \textrm{11$_g$}
\end{tabular}
\end{ruledtabular}
\end{table}

\begin{figure*}
\includegraphics[width=0.98\textwidth]{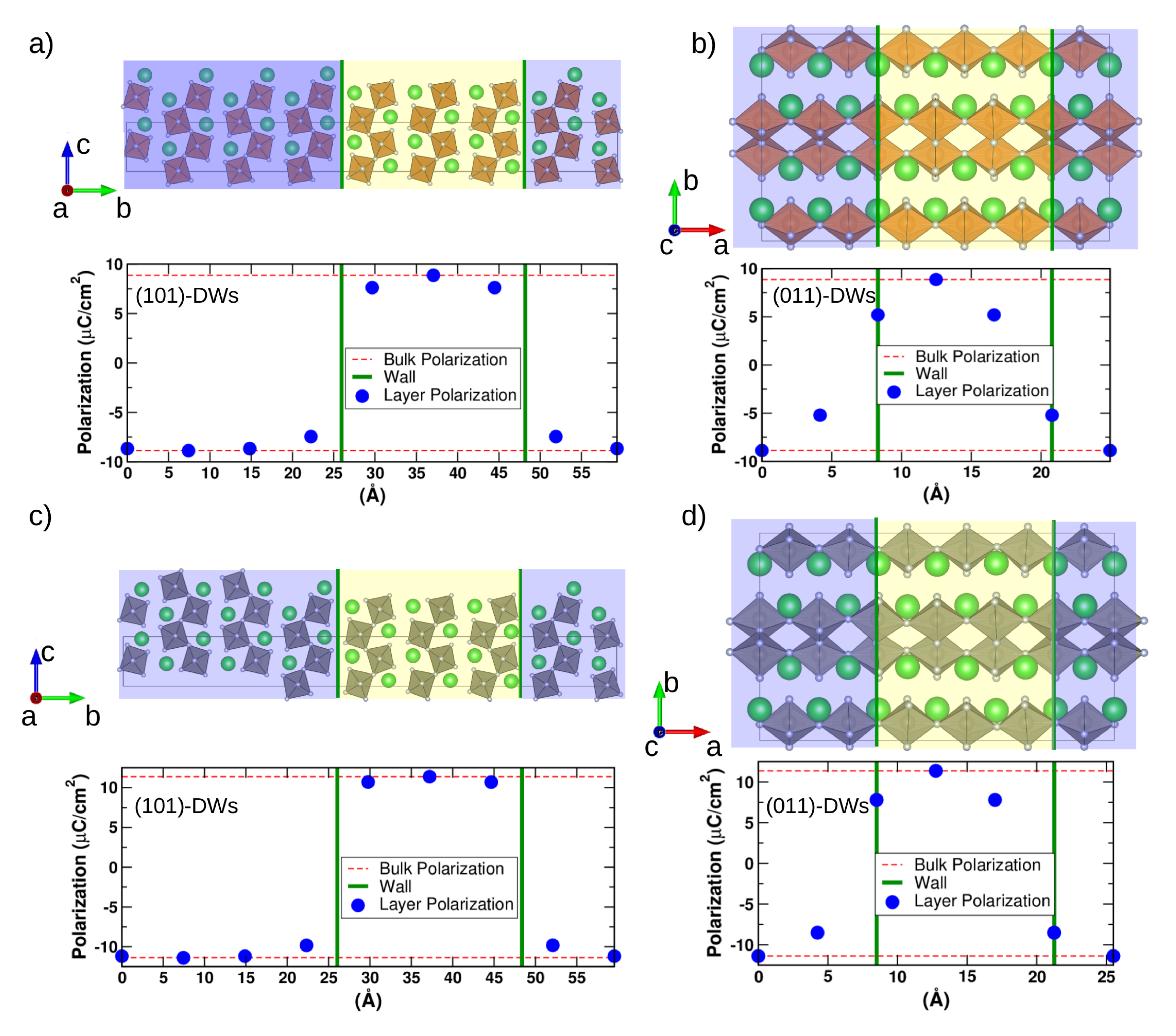}
\caption{\label{ac_bc_DW_struc_polar} (Color online) Layer Polarization in the (101)-DW and (011)-DW configurations for a,c) BaMgF$_4$ and b,d) BaZnF$_4$, respectively.}
\end{figure*}

We see that the domain wall energies of the fluorides are similar to those of the oxides but we find no clear correlation between 
Curie temperature, magnitude of ferroelectric polarization, and domain wall energies for either the conventional ferroelectric 
perovskites or the geometric ferroelectric compounds. 

We find that the domain wall energy is lowest for the (101)-DW configuration in BaMgF$_4$ and for the (011)-DW configuration in BaZnF$_4$
(although in the latter case the energies of the two wall types are very close). We attribute this difference to the chemical activity
of the Zn $3d$ electrons, although a detailed explanation is still lacking. 

In Fig. \ref{ac_bc_DW_struc_polar} we show our calculated layer-by-layer polarizations perpendicular to the wall direction
obtained from 
\begin{equation}
P=\frac{e}{\Omega}\sum_\alpha{Z_\alpha\cdot u_\alpha}. 
\end{equation}
Here $e$ is the charge of the electron, $\Omega$ the volume of a 1x1x1-cell layer, 
$u_\alpha$ is the displacement of atom $\alpha$ from its paraelectric position in the $z$-direction, $Z_\alpha$ are the formal ionic charges
and the index $\alpha$ runs over all atoms in the considered 1x1x1-cell layer.
Figs. \ref{ac_bc_DW_struc_polar}a) and \ref{ac_bc_DW_struc_polar}b) show our results for  BaMgF$_4$ and Figs. \ref{ac_bc_DW_struc_polar}c) and 
\ref{ac_bc_DW_struc_polar}d) for  BaZnF$_4$. We find that the (101)-DW domain walls are sharp (Figs. \ref{ac_bc_DW_struc_polar}a) and c)), 
that is, the local polarization changes abruptly across the domain wall from one bulk 
value to the opposite value. In contrast, our calculated lowest energy (011)-DW walls (Figs. \ref{ac_bc_DW_struc_polar}b) and d)) show a smoother 
change in polarization across the wall. 
We can rationalize the difference in wall
widths from the connectivity of the layers: While in the (101)-DW structures the wall lies between planes of MF$_6$ octahedra,
in the (011)-DW the walls cut through the planes enforcing a gradual change of polarization across the wall. 
We note also that for both wall geometries, the MgF$_6$ octahedra retain their regular
shape, but the ZnF$_6$ octahedra distort across the wall, particularly in the (011)-DW structure. 

We point out that such atomically sharp domain walls are also observed in the improper ferroelectric hexagonal YMnO$_3$ series 
\cite{kumagai,zhang2012} where they have been compared to the narrow twin planes formed at anti-phase 
boundaries in antiferrodistortive materials. Here the physics is similar, although the BaMF$_4$ series represents an unusual example
of the behavior in a {\it proper} ferroelectric. In conventional ferroelectric perovskite oxides, while ferroelectric domain walls tend
to be narrow, they are not atomically sharp. 

\section{Summary}
In summary, using first-principles density functional theory we have calculated two properties of BaMgF$_4$ and BaZnF$_4$ 
that are relevant for their behavior in ferroelectric thin films. 
First, we calculated the strain dependence of the spontaneous polarization and found that it varies close to linearly with both
compressive and tensile strain indicating that it can be tuned in coherent thin-film heterostructures by appropriate choice of substrate
lattice constant. 
Next, we calculated the energies and structures of neutral 180$^\circ$ domain walls and identified those most likely to occur in 
practical samples. We found that the domain wall energies are comparable to those of conventional oxide ferroelectrics, 
but that the wall thicknesses are thinner, reminiscent of twin boundaries in antiferrodistortive materials. 
We hope that our study motivates further experimental investigation of the BaMF$_4$ class of materials, and other unconventional
ferroelectrics. 

\begin{acknowledgments}
This work was supported by ETH Z\"urich and by the ERC Advanced Grant program, No. 291151. Computations were performed at the Swiss
Supercomputing Center and on the ETH Brutus cluster; final computational details and manuscript writing were completed at MIPT. The 
authors thank J. F. Scott for useful discussions.
\end{acknowledgments}

\bibliography{BaMF4_paper}

\end{document}